\documentclass[manuscript, nonacm]{acmart} 

\copyrightyear{2026}
\acmYear{2026}
\setcopyright{cc}
\setcctype{by-nc-nd}

\renewcommand\footnotetextcopyrightpermission[1]{} 
\AtBeginDocument{%
  }

\usepackage[utf8]{inputenc}
\usepackage{booktabs} 
\usepackage{tabularx}
\usepackage{graphicx}
\usepackage{amsthm}
\usepackage{float}

\newtheorem{hyp}{Hypothesis}
\definecolor{BabyBlue}{rgb}{0.54,0.81,0.94}
\definecolor{VividOrange}{rgb}{1.0, 0.5, 0.0}

\begin{document}

\title{Let Me Introduce You: Stimulating Taste-Broadening Serendipity Through Song Introductions}

\author{Brett Binst}
\email{brett.binst@vub.be}
\orcid{0000-0002-0243-9952}
\affiliation{%
  \institution{imec-SMIT, Vrije Universiteit Brussel}
  \city{Brussel}
  \country{Belgium}
}

\author{Ulysse Maes}
\email{ulysse.jan.l.maes@vub.be}
\orcid{0009-0005-6823-0237}
\affiliation{%
  \institution{imec-SMIT, Vrije Universiteit Brussel}
  \city{Brussel}
  \country{Belgium}
}

\author{Martijn C. Willemsen}
\email{m.c.willemsen@tue.nl}
\orcid{0000-0001-5908-9511}
\affiliation{%
  \institution{Eindhoven University of Technology}
  \city{Eindhoven}
  \country{The Netherlands}
}
\affiliation{%
  \institution{Jheronimus Academy of Data Science}
  \city{’s-Hertogenbosch}
  \country{The Netherlands}
}

\author{Annelien Smets}
\email{annelien.smets@vub.be}
\orcid{0000-0003-4771-7159}
\affiliation{%
  \institution{imec-SMIT, Vrije Universiteit Brussel}
  \city{Brussel}
  \country{Belgium}
}


\begin{abstract}
Research on how people experience music emphasizes the importance of exploration and diversity in listening.
However, music recommender systems struggle with facilitating exploration.
Even when music recommender systems are able to recommend something valuable to users that is outside their typical preferences, it still remains difficult to spark their interest.
This paper presents a user study examining the efficacy of immersive and informative introductions in stimulating interest in songs that are beyond one's usual preferences, an experience called Taste-Broadening Serendipity.
We uncover two important mechanisms behind the effect of introductions: transportation and cognitive elaboration.
Our findings indicate that transportation (i.e., being absorbed into a narrative world) is the strongest predictor of Taste-Broadening Serendipity, while cognitive elaboration (i.e., learning something new about the artist or social context in which the music emerged) has a weaker effect but is easier to stimulate.
We propose that song introductions can play an important role in facilitating exploration and increasing diversity of listening on music streaming platforms.
\end{abstract}



\maketitle

\section{Introduction}
In 2015, there were 67 million global paid subscribers of music streaming services.
By the end of 2024, this number rose to over 800 million~\cite{mulligan_music_2025}.
The appeal of music streaming services lies in their promise: a virtually limitless catalogue of music available in one's pocket, anytime and anywhere.
This abundance creates unprecedented opportunities for discovering new music.
To help listeners navigate this vast ocean of songs, platforms rely on Music Recommender Systems (MRS). 

\citet{anderson_algorithmic_2020} have found that exploration and diversity of listening are important predictors for long-term user conversion and retention.
This makes sense since listeners like to keep music listening fresh, seeking ``exposure in ways that open them up to new and unfamiliar styles and pieces, ensuring that their preferred music does not become disliked'' \cite{greasley_keeping_2016}.
However, MRS struggle with supporting users in exploring music outside their typical preferences \cite{anderson_algorithmic_2020, shakespeare_reframing_2025}.
The challenge is to recommend something outside the user's profile that also sparks their interest \cite{liang_personalized_2019}, an experience we call ``Taste-Broadening Serendipity'' \cite{binst_what_2025}.

To address this challenge, we draw inspiration from the psychology of interest.
According to \citeauthor{silvia_interestcurious_2008}'s \cite{silvia_interestcurious_2008} appraisal theory on interest ``new and comprehensible things are interesting; new and incomprehensible things are confusing.'' 
In other words, \citet{silvia_interestcurious_2008} states that a person experiences interest when a stimulus is perceived as complicated yet within one's coping potential. 

Previous research has manipulated coping potential and perceived complexity of content by providing introductions \cite{darda_impact_2023, silvia_what_2005}. 
However, the driving mechanisms behind these introductions are poorly understood, making it unclear why some introductions work and others do not \cite{hellmuth_margulis_when_2010}.

In this study, we explore two possible mechanisms to increase perceived complexity and coping potential when discovering unfamiliar music: Transportation and Cognitive Elaboration.
Transportation is a convergent process, where all attention is focused on events occurring in a narrative world \cite{green_role_2000} and is associated with positive effects on interest in, and engagement with content \cite{dahlstrom_using_2014}.
Cognitive elaboration, on the other hand, is a learning process that entails connecting new information to already known information, such as prior knowledge or personal experience. 
Cognitive elaboration has been found to facilitate learning of new information \cite{eveland_cognitive_2001, frauhammer_how_2025}.

Based on previous research, we expect these mechanisms to be triggered through different types of introductions: while transportation requires immersive content containing a protagonist, rich description of scenery, and a narrative arc \cite{green_role_2000}, cognitive elaboration requires establishing new associations \cite{eveland_cognitive_2001}.
Accordingly, this paper explores whether we can trigger transportation and cognitive elaboration through, respectively, immersive and informative introductions; whether these psychological mechanisms influence perceived complexity and coping potential; and whether these two pathways facilitate Taste-Broadening Serendipity when listeners discover unfamiliar music. 

\section{Related Literature}
\subsection{Taste-Broadening Serendipity in Music Recommender Systems}
While scholars agree on the importance of designing for serendipity in Recommender Systems, there is less agreement on what serendipity exactly entails \cite{kotkov_survey_2016,ziarani_serendipity_2021}.
To reduce this conceptual ambiguity, \citet{binst_what_2025} propose a unified framework centered on experienced serendipity, that is serendipity as a user experience \cite{smets_intended_2025}, where experienced serendipity is defined as ``fortuitously encountering content that is perceived as both refreshing and enriching.''
Importantly, the framework differentiates multiple ``flavors'' of experienced serendipity, capturing distinct ways in which serendipity can be experienced in practice \cite{binst_what_2025}.
These flavors more explicitly capture why an encounter feels fortuitous, refreshing, or enriching; for instance, because it is highly relevant to a current need, or because it is intriguing and sparks new interests \cite{binst_what_2025}.

In this study, we focus on one of the identified flavors, namely Taste-Broadening Serendipity defined as the experience of encountering items that are unusual and, at the same time, intriguing/interesting \cite{binst_what_2025}.
Several studies have been conducted on stimulating taste-broadening through MRS. \citet{liang_personalized_2019} distinguish between two approaches: pathway-based recommendation towards a target genre, and direct recommendation of songs belonging to a target genre.

\subsubsection{Pathway-Based MRS} 

Pathway-based MRS aim to recommend the smoothest path towards a target genre. For example, \citet{taramigkou_escape_2013} developed an MRS that stretches user preferences by calculating the optimal path from user preference graphs towards a selected latent genre. 
Similarly, \citet{flexer_playlist_2008} developed an approach to recommend a playlist that smoothly transitions between a chosen start and end song.
Lastly, \citet{nonnemaker_feedback-driven_2025} also studied taking gradual steps from current preferences towards a target genre, but additionally integrated user feedback through Bayesian active learning which further enhanced music exploration. 

\subsubsection{Direct Recommendation of a Target Genre}

Another approach to taste broadening is to expose users immediately to the target genre. 
Different ideas have been tested.

One approach is to visualize recommendations in the information space.
\citet{liang_interactive_2021} find that a contour plot visualizing the relationship of recommendations to the user profile and a target genre is perceived as helpful to explore a new genre. 

Another approach is through context-aware MRS. 
The underlying idea is that creating a context-song match will stimulate acceptance of unfamiliar songs.
\citet{abdollahpouri_towards_2018} find that the previous sequence of ads and songs has a strong influence on the success of exploration through an MRS.
However, apart from this study, limited research has been done to further explore this idea \cite{aman_interacting_2017}.

A third option is to provide users control over an MRS.
\citet{liang_interactive_2021} found that providing a slider to communicate mood, in combination with the contour plot discussed above, supports users in exploring the item space.
Providing mood control in MRS taps into the mood regulation function of music \cite{thoma_regulation_2006}.

A fourth approach is nudging, which posits that providing defaults towards exploration will result in more exploration.
\citet{liang_role_2021,liang_exploring_2022} indeed observed that users are more likely to discover distant genres if these were ranked highest in the genre selection list.
They also find that setting a genre-representativeness slider on ``representative'' by default, encourages users to explore songs that deviate from their current preferences \cite{liang_role_2021}.
However, in a longitudinal study, they observe that the long-term effect of nudging seems to be limited \cite{liang_exploring_2022}.

Lastly, various recommendation algorithms for exploration, different from the pathway approach discussed earlier, have been developed.
For instance, \citet{zhang_auralist_2012} proposed an algorithm that applies Latent Dirichlet Allocation to model listener communities associated with artists.
Next, they captured the listener diversity associated with artists and combined these signals with a declustering algorithm to recommend artists outside a user’s bubble \cite{zhang_auralist_2012}. \citet{mehrotra_algorithmic_2021} took a different approach by balancing three goals namely discovery, familiarity, and similarity.
He found that hierarchical ordered weighted averaging of these goals is superior over other tested techniques since it is able to improve both discovery and satisfaction over the baseline \cite{mehrotra_algorithmic_2021}.

However, none of these methods address the challenge of explicitly sparking users' \textit{interest} in the recommended items, which is a crucial aspect of achieving Taste-Broadening Serendipity.
Research on segues in music recommendation offers a promising, albeit underexplored, avenue for addressing this challenge \cite{behrooz2019augmenting,gabbolini2021generating}.
Segues are short informational bridges connecting consecutive songs, designed to enrich the listening experience by augmenting them with contextual background information \cite{behrooz2019augmenting}.
Prior work suggests they may be particularly effective in supporting music discovery \cite{behrooz2019augmenting}. 
However, a limitation of this work is that it looks at the connections between songs, rather than the potential serendipitous experience of listening to a specific song. 
So we will aim to augment songs through introductions rather than segues. 
Moreover, there is limited understanding on the psychological mechanisms behind effectively augmenting songs with background information. 
In the following sections, we draw on the psychology of interest to address this gap.

\subsection{The Psychology Behind Interest}
\citet{silvia_interestcurious_2008} states that something is experienced as interesting when it is perceived as both "complicated" yet within the coping potential of the person.
Complicated is understood broadly here and can mean “information rich” like Picasso’s Guernica, or contradicting a person’s beliefs, such as quantum physics for most people.
Coping potential refers to a person’s perceived ability to comprehend this complexity.

Silvia’s theory implies that we can stimulate interest by manipulating perceived complexity and coping potential.
Indeed, research supports this idea \cite{darda_impact_2023,silvia_cognitive_2005}.
For example, \citet{darda_impact_2023} find that providing contextual information about artworks in an exhibition increases both interest and liking, and \citet{silvia_cognitive_2005} finds that providing background information about poetry sparks interest.

In the context of Music Recommender Systems, \citet{maccatrozzo_everybody_2017,maccatrozzo_role_2023} have applied Silvia’s appraisal theory through the SIRUP model.
This model states that serendipity in MRS can be stimulated by leveraging two checks: a novelty check, i.e., whether the item is new compared to the user profile by using Linked Open Data paths, and a coping potential check, i.e., the user’s ability to cope with new content, operationalized as diversity of the user profile \cite{maccatrozzo_everybody_2017}.
In a later study, \citet{maccatrozzo_role_2023} then used the novelty check as a metric for evaluating experienced serendipity when using the system.
However, this approach risks conflating what \citet{smets_intended_2025} distinguishes as ``experienced serendipity,'' i.e., the user’s subjective experience of serendipity (for example, encountering something new and interesting) with ``afforded serendipity,'' which refers to the system features designed to facilitate such encounters, such as recommending more unusual items. In this case, \citet{maccatrozzo_role_2023} arguably evaluated afforded serendipity.

In this work, we use Silvia’s appraisal theory of interest \cite{silvia_interestcurious_2008} as a lens for designing afforded serendipity in the music domain. 
Although the theory characterizes interest in terms of two underlying appraisals (i.e., perceived complexity and coping potential), it is less explicit about how a system might reliably influence them. 
To address this gap, we examine two mechanisms that could help elicit these appraisals in response to recommended songs: transportation into the song’s narrative and cognitive elaboration of the song.
Furthermore, we draw on prior work on segues \cite{gabbolini2021generating,behrooz2019augmenting} to investigate a method for augmenting the listening experience in ways that may spark interest in recommended songs. 
However, instead of segues, we examine different types of introductions designed to stimulate either transportation or cognitive elaboration. 

\subsection{Two Mechanisms to Stimulate Taste-Broadening Serendipity}
\paragraph{Transportation} Transportation is defined as the experience of being absorbed in a narrative world \cite{green_narrative_2024} and is characterized by three main features: (1) cognitive processing of the story, (2) empathizing with the characters and imagining the scenery, and (3) loss of self-consciousness and track of reality \cite{van_laer_extended_2014}.

When listeners are transported into a narrative, they show less critical scrutiny \cite{green_role_2000}.
Suspending listeners’ judgement might make them more inclined to accept unfamiliar genres. 
Additionally, research finds that processing fluency, i.e., how fluently people process an artwork, increases aesthetic pleasure \cite{reber_processing_2004}.
Transportation stimulates processing fluency because the narrative organizes the chaos and inhibits the ambiguity, in this way effectively increasing coping potential \cite{dahlstrom_using_2014}.

Therefore, we hypothesize that transportation will increase listeners’ perceived coping potential and therefore the likelihood of experiencing Taste-Broadening Serendipity.
We will stimulate transportation through immersive introductions that detail a protagonist's story (see \autoref{fig:fig2} for an example).
We pose the following hypothesis on how immersive introductions can foster Taste-Broadening Serendipity:
\begin{hyp}
Immersive introductions foster Taste-Broadening Serendipity via a serial mediation path: increasing transportation, which in turn enhances coping potential.
\end{hyp}

\paragraph{Cognitive Elaboration}
\citet{bourdieu_distinction_1984} demonstrates that musical taste is not a natural gift but a product of socialization.
Enjoying music, especially less accessible music like Jazz and Opera, requires cultural capital, i.e., the background knowledge necessary to decipher the artwork \cite{bourdieu_distinction_1984}.
Without cultural capital, unfamiliar music often appears as chaotic or meaningless.

To overcome this barrier, we turn to \citet{leder_model_2004} their model of art consumption as “cognitive mastering.” This considers listening as a problem-solving task where the observer attempts to find a satisfactory interpretation of the artwork.
If cognitive mastering is successful, the ambiguity of the music is resolved and the music is perceived as comprehensible, leading to a rewarding sense of "aesthetic insight" or understanding.

In contrast to transportation, which operates through emotional immersion and a "loss of self," cognitive mastering requires a more rational, analytic mode of processing.
We expect to stimulate this understanding through cognitive elaboration, defined as the active process of connecting new information to prior knowledge \cite{eveland_cognitive_2001}.
We posit that informative introductions can provide "situational cultural capital,” effectively giving the listener the keys to solve the musical puzzle.
Therefore, we attempt to stimulate cognitive elaboration through informative introductions that uncover the social context and inspiration behind the song (see \autoref{fig:fig2}).
We formulate the following hypothesis:

\begin{hyp}
Informative introductions foster Taste-Broadening Serendipity via a serial mediation path: increasing cognitive elaboration, which subsequently increases both coping potential and perceived complexity.
\end{hyp}

\begin{figure}[H]
    \centering    \includegraphics[width=0.8\linewidth]{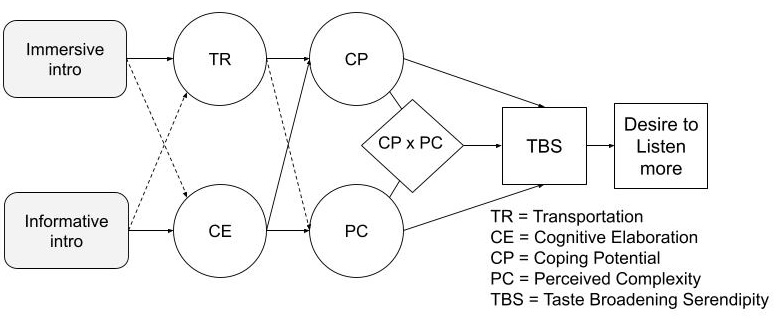}
    \caption{Path diagram illustrating the two pathways hypothesized to stimulate Taste-Broadening Serendipity.}
    \label{fig:fig1}
\end{figure}

\autoref{fig:fig1} displays the conceptual model, following the user-centric evaluation framework of \citet{knijnenburg_evaluating_2015}.
The model connects users’ perceptions of transportation (TR) and cognitive engagement (CE) to coping potential (CP) and perceived complexity (PC) as drivers of Taste-Broadening Serendipity (TBS).
The full arrows in~\autoref{fig:fig1} summarize the two hypothesized pathways to stimulate Taste-Broadening Serendipity as discussed above.
Although we expect that immersive introductions (compared to no introduction) will mainly stimulate Taste-Broadening Serendipity through the serial mediation of transportation on perceived coping potential, and informative introductions will mainly work through the serial mediation of cognitive elaboration on perceived complexity and coping potential, we will also test the effect the introductions have on the other pathway (i.e., the dotted lines).
This way, we can explore whether they reinforce each other, are negatively correlated, or are completely independent.
Lastly, we will test whether the frequency of experiencing Taste-Broadening Serendipity in a session is predictive for the desire to listen to more songs from this genre as a proxy for actual taste-broadening.

\section{Methodology}
\subsection{Implementation}\label{implementation}
To test our hypotheses, we developed a custom web-based application which allowed a user study with remote, asynchronous participation across desktop and mobile platforms.
The application was hosted on Vercel (vercel.com), deploying a PostgreSQL database (neon.com) for GDPR-compliant data storage.
The system captured measures of user engagement (likes/dislikes, song skips, total listening time), and behavior (clickstream data and attention checks), as well as subjective measures of user experience.
The full code for the application can be found in a\href{https://github.com/imec-smit-vub-mep-rec/Music-Introductions-Experimental-Environment.git}{\textit{ Github repository}}\footnote{https://github.com/imec-smit-vub-mep-rec/Music-Introductions-Experimental-Environment.git}.

To test the impact of different types of narratives on discovery, we curated a stimulus set of 6 genres, each containing 3 songs.
Genres were selected based on the taxonomy by \citet{rentfrow_re_2003} to ensure a diverse representation of four musical styles: reflective and complicated (Opera and Blues), intense and rebellious (Heavy Metal), upbeat and conventional (Country and Disco), and energetic and rhythmic (Reggae).
For each of the six genres, we selected three songs based on their representation of the genre and narrative potential (e.g., ``Smalltown Boy'' for Disco, ``Habanera'' for Opera).
For each song, we authored three introduction styles: Informative (biographical/factual), Immersive (narrative-driven) and Control (no introduction).
These introductions were synthesized using Google Vertex AI Studio (Gemini 2.5 Text-To-Speech (TTS) model~\footnote{https://blog.google/innovation-and-ai/technology/developers-tools/gemini-2-5-text-to-speech/)}).
We applied genre-specific vocal prompts and adjusted the speech rate to 1.25x to make the voice sound natural.

\begin{figure}[H]
    \centering
    \includegraphics[width=0.95\linewidth]{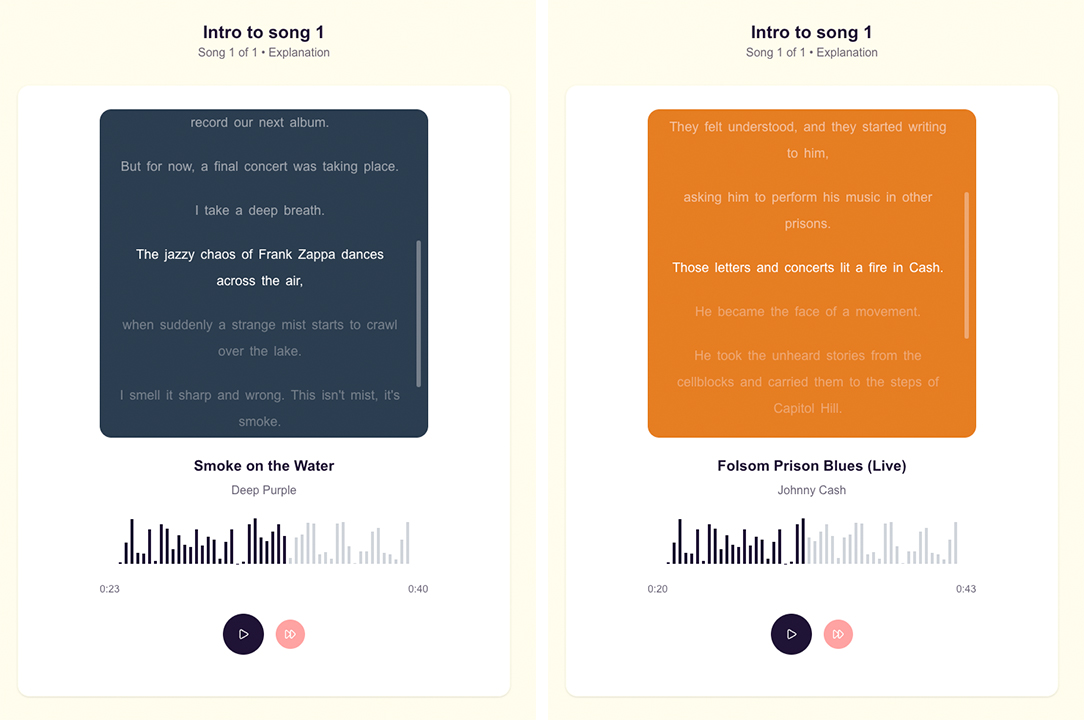}
    \caption{Left: An immersive intro for “Smoke on the Water”.
    Right: an informative intro for “Folsom Prison Blues”.}
    \label{fig:fig2}
\end{figure}

\subsection{Experimental Design}
Each participant completed three trials in which each introduction style (Control, Informative, or Immersive) was tested within-subjects and counterbalanced to control for order effects.
{By randomly assigning songs and introduction types to participants we neutralize systematic bias across our experimental conditions. 
Furthermore, due to our multilevel statistical analyses and within-subject design, we control for differences between participants (e.g., familiarity with the genre).}
Lastly, as this paper focuses purely on Taste-Broadening Serendipity, we ensure the validity of our focus on songs outside the user’s typical preferences by including only those trials in which songs were confirmed to be unfamiliar, based on participants’ agreement with the item ``The song differs from the type of music I normally listen to''.
The experimental procedure was structured into four phases:
\begin{enumerate}
    \item \textbf{Onboarding}: Upon informed consent, participants provided basic demographic information (country and date of birth).
    \item \textbf{Listening Task}: Participants listened to three distinct songs.
    While each participant listening to the same genre heard the same three songs, both the order of the songs and their accompanying introduction style (Informative, Immersive, or No Introduction) varied across sessions.
    Each of the three song trials consisted of an audio component followed by a survey.
    The audio player featured genre-specific visual cues, time-synced lyrics during spoken introductions, and an animated progress bar (see \autoref{fig:fig2}).
    \item \textbf{Post-Listening Measures}: Immediately after listening to each song, participants rated their experience on six dimensions (Taste-Broadening Serendipity, transportation, cognitive elaboration, coping potential, perceived complexity, quality of introduction) using 5-point Likert scales.
    \item \textbf{Post-Experiment Measures}: After having listened and answered questions about 3 songs with their accompanying introductions, a final questionnaire was shown to measure overall satisfaction and desire to listen to more songs from the explored genre.
    We also provided an open text option for participants to share their thoughts about the study.
\end{enumerate}

\subsection{Recruitment and Participants}
The experiment was deployed in phases, beginning with a pilot on October 29, 2025, followed by a full release among our own network on November 20 as a safety check for application stability, before starting the Prolific deployment on December 2.
Data collection concluded on December 8.
We recruited 475 participants (402 via Prolific, 73 via snowball sampling).
We actively sampled for diversity in education level on Prolific through stratified sampling.
Strict exclusion criteria were applied to ensure data quality: sessions were rejected if participants failed two or more attention checks or provided incomplete responses.
Throughout the study, we included three attention checks (e.g.,  ``Indicate `Strongly disagree' to continue'').
After filtering for the unfamiliar condition---validated by the item ``The song differs from the type of music I normally listen to''---we retained 350 relevant sessions/participants (Age: Mean = 36, SD = 12.79; Gender: Female = 48.29\%, Male = 50.29\%, Other = 1.42\%).
The sample spanned 37 countries, with the highest representation from South Africa (n=54), UK (n=50), and Belgium (n=33).

\subsection{Measurement}

\begin{table}[H]
    \centering
    \caption{Reliability of the scales used to measure subjective user perceptions and experiences after listening to a song.}
    \fontsize{8}{10}\selectfont 
    \renewcommand{\arraystretch}{0.95}
    \label{tab:tab0}
    \footnotesize
    \begin{tabularx}{\linewidth}{@{}Xr@{}} 
        \toprule
        \textbf{Scale and Items} & \textbf{Loading} \\
        \midrule
        \textbf{Cognitive Elaboration ($H = .94$)} & \\
        I gained new insights about the song or artist & 0.872 \\
        I reflected on how the song connects to broader ideas or culture & 0.732 \\
        I learned something new about the song or artist I listened to & 0.945 \\
        I learned more about the background behind the song or artist & 0.905 \\
        \addlinespace
        
        \textbf{Coping Potential ($H = .94$)} & \\
        I feel like I get the song & 0.887 \\
        The song makes sense to me & 0.908 \\
        I understand the song & 0.889 \\
        I feel like I get what the song is about & 0.884 \\
        I find the song confusing & 0.748 \\
        \addlinespace
        
        \textbf{Perceived Complexity ($H = .90$)} & \\
        The song expresses a deep meaning & 0.685 \\
        This song feels superficial (without depth) to me & 0.808 \\
        There isn’t much depth to this song & 0.931 \\
        \addlinespace
        
        \textbf{Transportation} \cite{appel_transportation_2015} ($H = .85$) & \\
        I could picture myself in the scenes of the song & 0.792 \\
        I was mentally involved in the song while listening & 0.821 \\
        The song affected me emotionally & 0.657 \\
        While listening to the song I had a vivid image of what it was about & 0.729 \\
        \bottomrule
        \addlinespace
        \multicolumn{2}{@{}p{\linewidth}@{}}{\footnotesize \textit{Note:} After each item we communicate its standardized factor loading on the latent variable. $H$ represents the reliability coefficient for the scale.}
    \end{tabularx}

\end{table}

In~\autoref{tab:tab0}, we list the scales presented to the participants after each song (except for the Taste-Broadening Serendipity scale which we describe below) and their corresponding reliability scores.
All scale items are measured on 5-point Likert scales (strongly disagree to strongly agree).
The transportation scale was adopted from \citet{appel_transportation_2015}, but we dropped the item ``I wanted to learn how the narrative ended'' since it is less applicable in the context of music. 
The other scales were developed by the authors for this study. 

For the Taste-Broadening Serendipity scale, internal reliability is not relevant since it is a composite concept.
We operationalize\footnote{We also explored a fuzzy logic operationalization that maintained the same dimension thresholds ($>3$) but assigned weighted values to scores of 4 and 5 to create a continuous variable. This approach was discarded because the resulting distribution was difficult to model.} Taste-Broadening Serendipity, following \citet{binst_what_2025}, as scoring 4 or higher on item 1: “The song differs from the type of music I normally listen to” (note that this is always the case in our filtered dataset), scoring 4 or higher on item 2: “It sparked my interest,” and scoring 4 or higher on one of the other items (“I felt surprised,” “I discovered something unexpected,” “I discovered something that might have been more difficult to discover without the help of this app,” “I experienced the song through a new perspective,” and “I noticed details I didn’t notice before”).
The result is a binary variable with score 1 if all the above criteria are fulfilled and 0 otherwise.

\subsection{Analysis}
To estimate the mediation effects of our two theoretical pathways illustrated in~\autoref{fig:fig1}, we use multilevel Structural Equation Modeling (SEM).
As noted by \citet{knijnenburg_evaluating_2015}, SEM is a robust statistical framework to estimate latent constructs and all hypothesized paths in a single model.
This approach allows to explicitly model mediation of causal effects and correct for measurement error of latent constructs \cite{knijnenburg_evaluating_2015}.

We use Lavaan in R to perform the SEM analysis \cite{lavaan}.
We use the Maximum Likelihood Robust (MLR) estimator because it is easy to interpret, provides robust standard errors, and takes into account clustering in data.
Although the effect on the binary variable Taste-Broadening Serendipity is better modelled using the Weighted Least Squares Mean and Variance adjusted (WLSMV) estimator, it is not possible to model clustering using this estimator.
However, as a robustness check we do run the WLSMV model as well as a Bayesian multivariate regression model using brms \cite{burkner_brms_2017}.
Our analysis results are available in a~\href{https://github.com/imec-smit-vub-mep-rec/music-discovery-study-analysis.git}{Github repository}\footnote{https://github.com/imec-smit-vub-mep-rec/music-discovery-study-analysis.git}.

\section{Results}
\subsection{Introductions are Effective in Stimulating Taste-Broadening Serendipity}
If a song was recommended without an introduction, 38.4\% of the listeners reported experiencing Taste-Broadening Serendipity.
In contrast, this increased to 49.1\% if the song was preceded by an immersive introduction and even to 55.9\% in cases where the song was preceded by an informative introduction.

To test the effects of these introductions, we ran a Bayesian multilevel logistic regression.
\autoref{tab:tab1} illustrates that the positive effect of both introduction types (compared to the control condition) is significant.
The random effects suggest that there is quite some variation in the probability of experiencing Taste-Broadening Serendipity across participants, but also across songs.

\begin{table}[H]
    \centering
    \caption{\textbf{Effect of introduction type on experiencing Taste-Broadening Serendipity.}}
    \fontsize{8}{10}\selectfont 
    \renewcommand{\arraystretch}{0.95}
    \label{tab:tab1}
    \begin{tabular}{lcc}
        \toprule
        \textbf{Parameter} & \textbf{OR / SD\textsuperscript{1}} & \textbf{95\% Credible Intvl.} \\
        \midrule
        
        \textbf{Fixed Effects} & & \\
        \addlinespace[0.1cm] 
        Baseline (No Introduction) & 0.54 & (0.34 -- 0.84) \\
        Immersive Introduction & 1.93 & (1.24 -- 2.98) \\
        Informative Introduction & 2.66 & (1.74 -- 4.19) \\
        
        \addlinespace[0.3cm] 
        \textbf{Random Effects} & & \\
        \addlinespace[0.1cm]
        Song Variability (SD) & 0.54 & (0.28 -- 0.92) \\
        Participant Variability (SD) & 1.50 & (1.12 -- 1.92) \\
        \bottomrule
        
        \multicolumn{3}{l}{\footnotesize \textsuperscript{1} Fixed effects reported as Odds Ratios. Random effects reported as Standard Deviations.} \\
        \multicolumn{3}{l}{\footnotesize Estimates are posterior means.} \\
    \end{tabular}
\end{table}

\begin{figure}
    \centering
    \includegraphics[width= 0.99 \linewidth]{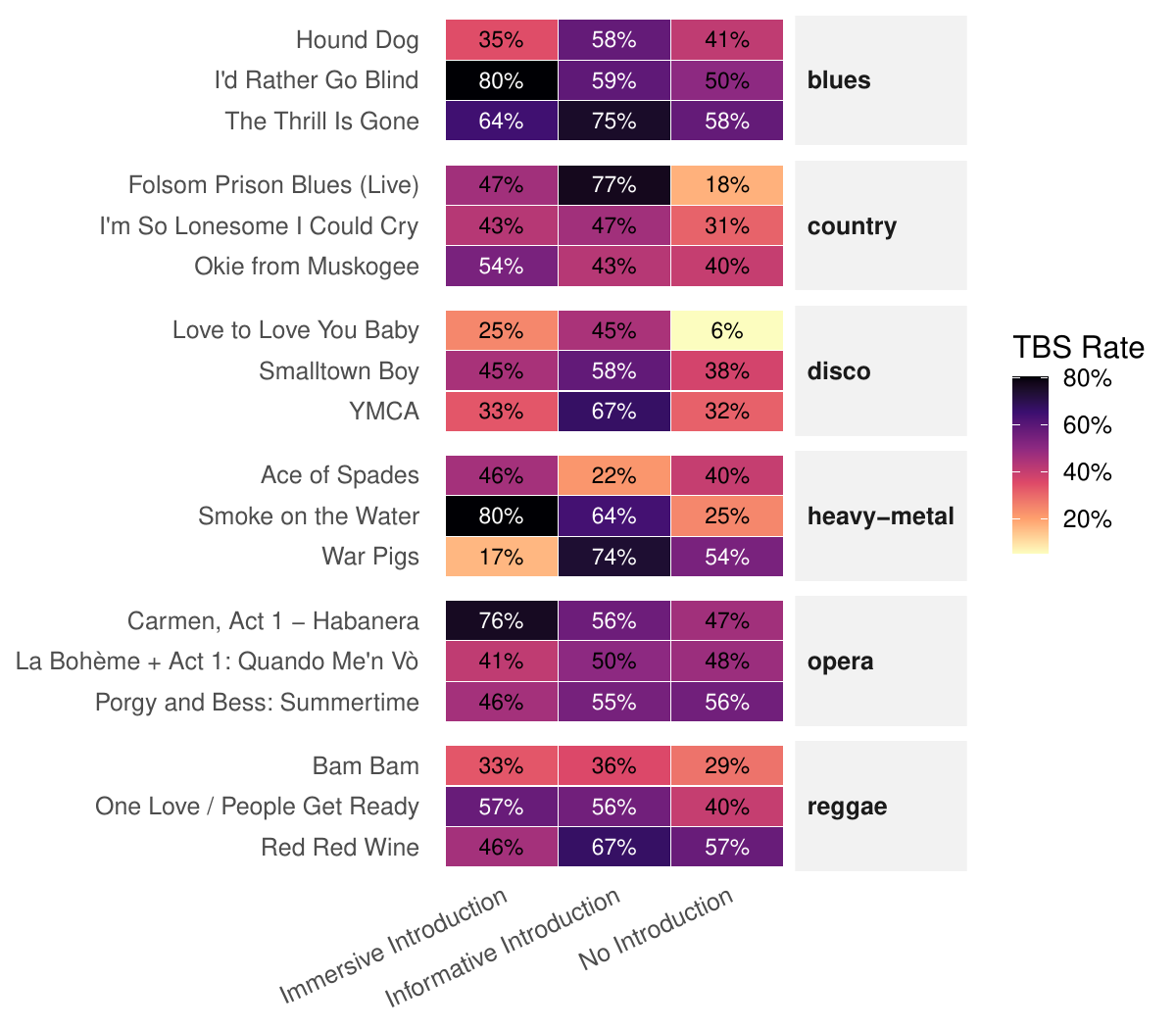}
    \caption{TBS rates by song and intro method.}
    \label{fig:fig3}
\end{figure}

\autoref{fig:fig3} illustrates the variation of experiencing Taste-Broadening Serendipity across songs.
For some songs, the introductions had a strong effect.
For example, 25\% experienced Taste-Broadening Serendipity after listening to ``Smoke on the Water'' without an introduction, while this increased to 64\% when it was preceded by the informative introduction, and 80\% when it was preceded by the immersive introduction.
For other songs, the introductions did not work or even had a negative effect, for example, the immersive introduction preceding ``War Pigs.''

Part of the explanation why the effects differ across songs is how interesting the introduction to the particular song is perceived to be.
To test this, we split the data based on the perceived interestingness of the introduction (measured on a 5-point Likert scale) into a low (< 3), medium (2-4), and high (> 3) condition.
We find that informative introductions that were not perceived as interesting actually have a negative effect on Taste-Broadening Serendipity compared to no introduction (for low quality immersive introductions, there was no significant difference).
Conversely, when introductions were perceived as highly interesting, they had a strong positive effect compared to the no introduction condition (immersive introductions: OR = 2.55, CI = 1.60–4.14; informative introductions: OR = 3.68, CI = 2.31–5.97).

\subsection{Why are Introductions Effective in Stimulating Taste-Broadening Serendipity?}
\subsubsection{Introductions Increase Perceived Complexity and Coping Potential}
\autoref{tab:tab2} illustrates that experienced Taste-Broadening Serendipity (TBS) is highest when both Coping Potential (CP) and Perceived Complexity (PC) are high. 

\begin{table}[ht]
    \centering
    \fontsize{8}{10}\selectfont 
    \renewcommand{\arraystretch}{0.95}
    \caption{\textbf{TBS by category}}
    \label{tab:tab2}
    
    \begin{tabular}{l|ccc}
        \hline
        \multicolumn{1}{c}{} & \multicolumn{3}{c}{\textbf{CP Category}} \\
        \cline{2-4} 
        
        \textbf{PC Category} & Low & Average & High \\
        \hline
        
        Low & 13.0\% & 23.0\% & 46.3\% \\
        \hline
        Average & 23.2\% & 55.2\% & 63.2\% \\
        \hline
        High & 47.8\% & 64.8\% & 87.2\% \\
        \hline
    \end{tabular}
\end{table}

\begin{figure}[H]
    \centering
    \includegraphics[width=1\linewidth]{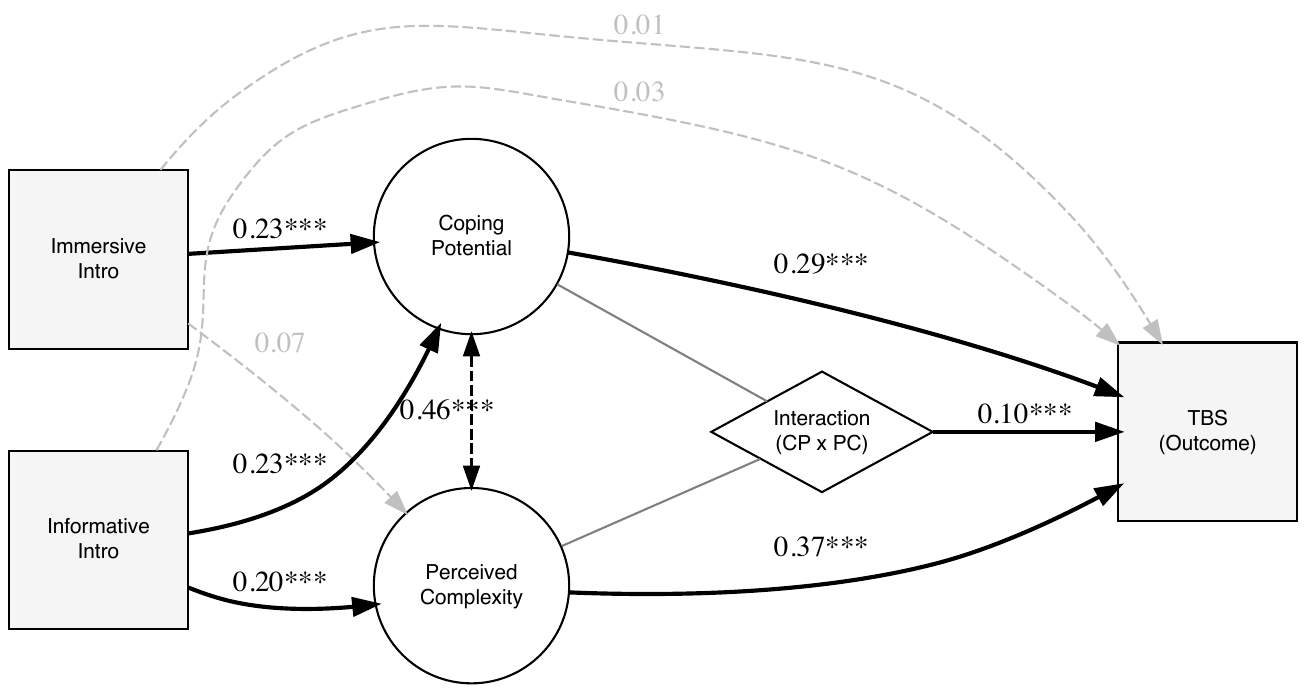}
    \caption{Path diagram of standardized (latent and observed) MLR estimates of the tested model. Constructs: latent (circles), observed (squares), interactions (diamonds).}
    \label{fig:fig4}
\end{figure}

To test whether the effect of our introductions on experiencing Taste-Broadening Serendipity can be explained by their influence on perceived complexity and coping potential, we used SEM.
The model fit was adequate (CFI = 0.954, TLI = 0.934, RMSEA = 0.082 \& SRMR = 0.034) so the estimates are trustworthy.
~\autoref{fig:fig4} illustrates that the effect of the immersive introductions on Taste-Broadening Serendipity is fully mediated by coping potential, and the effect of informative introductions on Taste-Broadening Serendipity is fully mediated by both perceived complexity and coping potential.
The robustness check (WLSMV estimator and multivariate Bayesian regression) confirms the full mediation of the introduction effects by perceived complexity and coping potential.

\subsubsection{Introductions Stimulate Transportation and Cognitive Elaboration}

\begin{figure}
    \centering
    \includegraphics[width=1.05\linewidth]{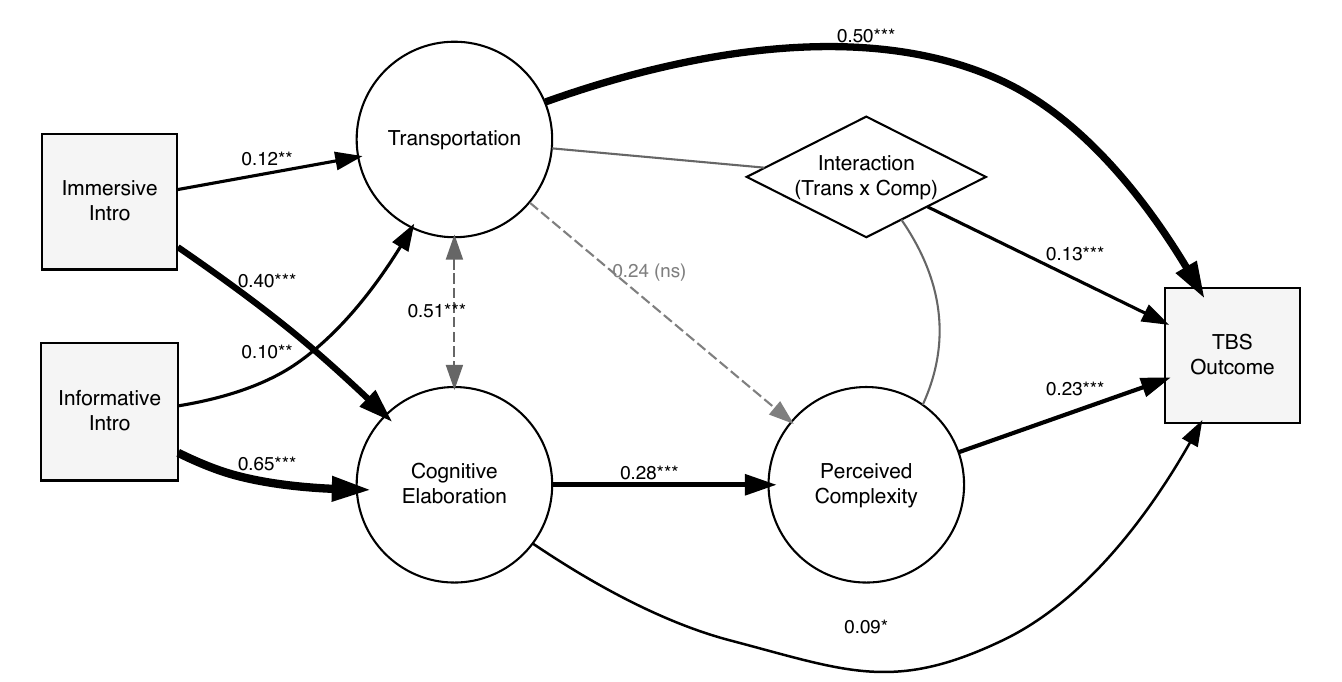}
    \caption{Path diagram of standardized (both latent and observed) estimates of the tested model through SEM. Latent constructs are represented by circles, observed constructs by squares, and interactions by diamonds.}
    \label{fig:fig5}
\end{figure}

While transportation and cognitive elaboration both increase coping potential and perceived complexity, they also directly impact Taste-Broadening Serendipity.
Furthermore, SEM analysis shows that the direct effect of transportation on Taste-Broadening Serendipity makes perceived coping potential redundant.
This indicates that transportation is the primary mechanism, and high coping potential is merely a byproduct of high transportation, not a separate driver of Taste-Broadening Serendipity.

\autoref{fig:fig5} illustrates our final model (i.e., without coping potential and with direct effects of both cognitive elaboration  and transportation).
The model has an adequate fit (CFI = 0.948, TLI = 0.932, RMSEA = 0.073 \& SRMR = 0.057), indicating that the estimates are trustworthy.
The robustness checks we ran (again using the WLSMV estimator in Lavaan and running a multivariate regression analysis in brms) confirm the significance of these estimates.

Transportation is the strongest predictor of experiencing Taste-Broadening Serendipity, followed by perceived complexity.
There is also a significant interaction between both, indicating that Taste-Broadening Serendipity is most likely to occur when listeners experience transportation in a song that is perceived as deep and complicated.
Lastly, we also observe a small direct effect of cognitive elaboration on Taste-Broadening Serendipity.

\begin{table}
    \centering
    \caption{\textbf{Indirect effects of introductions on TBS.} analysis of mediation pathways via TR, CE, and PC.}
    \fontsize{8}{10}\selectfont 
    \renewcommand{\arraystretch}{0.95}
    \label{tab:tab3}
    \resizebox{\columnwidth}{!}{%
        \begin{tabular}{lcccc}
            \toprule
            \textbf{Pathway} & \textbf{Unstd. Est.} & \textbf{SE} & \textbf{p-value} & \textbf{Std. Beta} \\
            \midrule
            
            \multicolumn{5}{l}{\textbf{Immersive Intros}} \\
            \addlinespace[0.1cm]
            Direct via TR & 0.063 & 0.021 & \textbf{0.003} & 0.060 \\
            Via TR and PC & 0.007 & 0.004 & 0.074 & 0.006 \\
            Via CE and PC             & 0.027 & 0.008 & \textbf{0.001} & 0.025 \\
            Direct via CE             & 0.037 & 0.016 & \textbf{0.025} & 0.034 \\
            \cmidrule(lr){1-5} 
            Total Effect on TBS       & 0.133 & 0.029 & \textbf{0.000} & 0.126 \\
            \cmidrule(lr){1-5} 
            
            \addlinespace[0.4cm] 
            
            \multicolumn{5}{l}{\textbf{Informative Intros}} \\
            \addlinespace[0.1cm]
            Direct via TR & 0.050 & 0.020 & \textbf{0.011} & 0.048 \\
            Via TR and PC & 0.005 & 0.004 & 0.123 & 0.005 \\
            Via CE and PC             & 0.042 & 0.013 & \textbf{0.001} & 0.040 \\
            Direct via CE             & 0.058 & 0.026 & \textbf{0.026} & 0.055 \\
            \cmidrule(lr){1-5} 
            Total Effect on TBS       & 0.156 & 0.033 & \textbf{0.000} & 0.148 \\
            
            \bottomrule
            \addlinespace[0.1cm]
            \multicolumn{5}{p{\linewidth}}{\footnotesize \textit{Note:} TR = transportation, PC = perceived complexity and CE = Cognitive Elaboration.} \\
        \end{tabular}
    } 
\end{table}

\autoref{tab:tab3} reports the two pathways towards Taste-Broadening Serendipity; through transportation and through cognitive elaboration.
We find that immersive introductions' most important pathway for increasing Taste-Broadening Serendipity is via transportation while informative introductions mainly influence Taste-Broadening Serendipity by stimulating cognitive elaboration.
However, in contrast to what we expected, the introductions actually work to a significant degree through both mechanisms.
Furthermore, transportation and cognitive elaboration are strongly correlated (r = 0.51).

\subsection{Taste-Broadening Serendipity Predicts the Desire for Genre Exploration}
\begin{figure}
    \centering
    \includegraphics[width=0.54\linewidth]{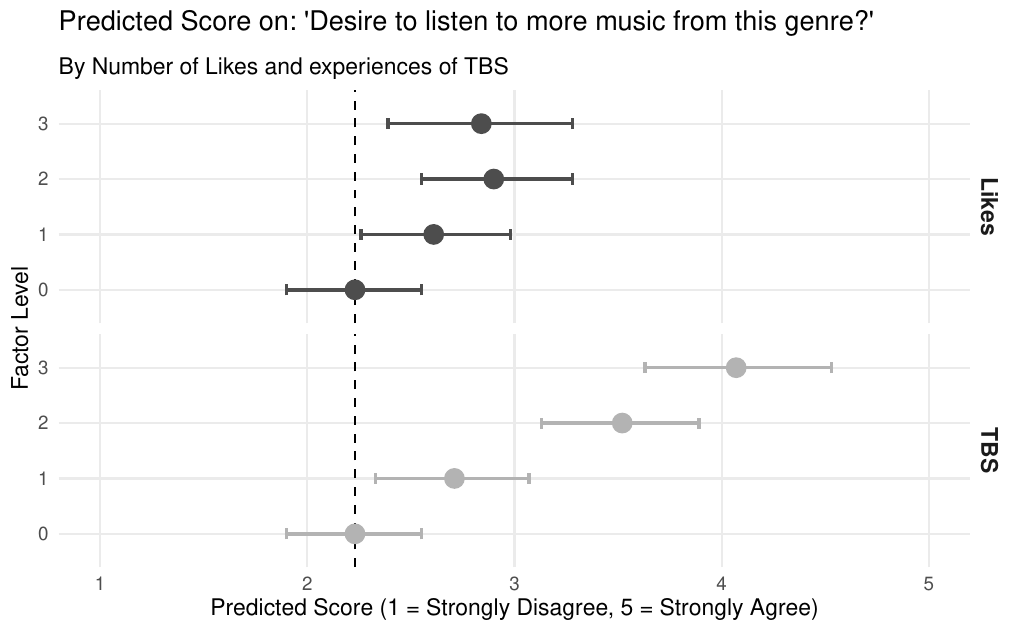}
    \caption{Predicted score on ``Desire to listen to more music from this genre''.}
    \label{fig:fig6}
\end{figure}

To test whether experiencing Taste-Broadening Serendipity predicts a user's desire to explore a genre, we conducted a Bayesian regression analysis.
The dependent variable, agreement with the statement `I would like to listen to more music from this genre,'' was predicted using two independent variables: the total number of liked songs and the number of songs that triggered Taste-Broadening Serendipity during a session. 
\autoref{fig:fig6} illustrates that the more songs within a session trigger experiencing Taste-Broadening Serendipity, the stronger the desire to listen to more songs from the same genre. For comparison, we contrast this against the effect of liking songs during the study; this effect is a lot smaller and caps after liking two songs.

\section{Discussion}
\subsection{The Potential of Introductions for Music Streaming Platforms}
Music exploration and diversity of listening are important predictors of long-term user conversion and retention on music streaming platforms \cite{anderson_algorithmic_2020}.
However, current Music Recommender Systems (MRS) struggle to support exploration \cite{shakespeare_reframing_2025}.
Even if MRS recommend interesting songs outside of listeners’ typical preferences, it remains challenging to spark listeners’ interest in these songs.

We find that both immersive and informative introductions help resolve this challenge by stimulating Taste-Broadening Serendipity.
And, as Figure \ref{fig:fig6} illustrates, experiencing Taste-Broadening Serendipity stimulates the desire to explore new genres.
Participants also expressed strong enthusiasm for these spoken introductions, with one noting: ``I would love to have a music service with this on it, it's brilliant.
Having artists introduce certain songs. Sign me up!''

Therefore, we encourage music streaming platforms to leverage immersive and informative introductions in the form of ``taste-breaker'' playlists. 
These playlists would introduce users to distant genres or unfamiliar artists by augmenting the songs with informative and immersive introductions.
This idea is similar to \citeauthor{gabbolini_user-centered_2022}'s \cite{gabbolini_user-centered_2022} music tours, a list of songs that are connected through segues, which are particularly liked during active listening.

Platforms could algorithmically detect when a user is trapped in a ``boredom bubble'' based on repetitiveness of listening during recent session. 
Recommender systems could then proactively suggest these taste-breaker playlists to deliver a refreshing listening experience.

\subsection{Stimulating Taste-Broadening Serendipity through Transportation and Cognitive Elaboration}

While previous research already suggested that augmenting songs with background information can spark users' interest \cite{gabbolini2021generating,behrooz2019augmenting}, it focused on the connection between songs via segues instead of introductions. 
Moreover, it remained unclear how and why these enhancements work in improving the listening experience.

We demonstrate that spoken introductions can successfully stimulate Taste-Broadening Serendipity by increasing perceived complexity and coping potential, and that transportation and cognitive elaboration drive these effects. 
Furthermore, we find that spoken introductions can successfully manipulate these perceptions through two mechanisms: transportation and cognitive elaboration.
However, in contrast to our expectations, transportation and cognitive elaboration also have an independent direct effect on Taste-Broadening Serendipity.
In fact, feeling transported into a song proved even more critical than its perceived comprehensibility.
As one participant noted: “I like the introductions, which put me in the atmosphere of the song.” 
Hence, we find that introductions are most effective when they transport listeners into songs they perceive as complicated/deep.

While we expected based on previous research \cite{eveland_cognitive_2001,green_narrative_2024} that immersive introductions would stimulate transportation and informative introductions would stimulate cognitive elaboration, results indicate that both introduction types activate both mechanisms.
Furthermore, transportation and cognitive elaboration are strongly correlated implying that both mechanisms are more alike than initially expected.

Lastly, we find informative introductions to be a reliable method to increase Taste-Broadening Serendipity because they have a strong effect on cognitive elaboration.
In contrast, immersive introductions proved to be a riskier method to increase Taste-Broadening Serendipity because their effect is less consistent.
This inconsistency could be caused by their reliance on transportation, a fragile psychological state. 
If the writing is poor, the listener does not get "transported," and the introduction fails \cite{green_role_2000}, or worse, backfires \cite{hellmuth_margulis_when_2010}.
Therefore, informative introductions offer the most accessible path to increasing Taste-Broadening Serendipity.
On the other hand, immersive introductions remain interesting for songs with strong narratives that afford transportation.
We conclude that there is no "one-size-fits-all" approach; instead, the most effective strategy varies based on the unique qualities of the song.

\subsection{Limitations and Future Research}

Although in general the study shows strong and clear effects of the two types of introductions, we observed considerable variation in the effects of the introductions as shown in Figure \ref{fig:fig3}. 
Our findings suggest that this variation is partly explained by differences in perceived interestingness of the introductions.
Future research could explore which characteristics of introductions drive interest, transportation, and cognitive elaboration. 
What makes an introduction effective in stimulating Taste-Broadening Serendipity might also differ from person to person and depend on genre and context.
For example, \citet{greenberg_musical_2015} show that musical preference depends on whether you are more of a music empathizer (looking for emotional resonance in music) or systemizer (interested in the structure of songs).
For empathizers, immersive introductions and transportation might be more potent in stimulating Taste-Broadening Serendipity, while for systemizers perhaps informative introductions and cognitive elaboration might be more effective.
\citet{behrooz2019augmenting} suggest that personalizing segues could make them more effective. 
The relationship between personality, genre, context, and the effectiveness of introductions provides an important avenue for future research aimed at personalizing these introductions.

A second limitation relates to the experimental setup, specifically the use of GenAI TTS.
We observed a negative spillover effect where the AI voice undermined perceived content credibility for some users (e.g., one participant commented ``I assumed that because the voice was AI, the information was also collected and constructed by AI which I know is faulty.''). 
As \citet{behrooz2019augmenting} note, voice characteristics are a primary factor in the success of augmenting songs with spoken segues.
Consequently, future research could investigate how varying voice characteristics impacts transportation and cognitive elaboration.

Finally, our study reflects an inherent trade-off between experimental control and ecological validity.
The requirement to listen to introductions and complete survey questions inevitably made the listening experience less naturalistic. 
Although participants were allowed to skip songs after 30 seconds, this only partially approximated organic listening behavior.
Hence, future research could test the potential of introductions in a more naturalistic setting. 
Nevertheless, our experiment strongly suggests that introductions are a promising feature to integrate into music streaming platforms. 

\section{Conclusion}
Music Recommender Systems struggle in supporting exploration.
Although they might recommend potentially interesting and unfamiliar items to the user, the challenge is to actually spark their interest in these items.
This paper demonstrates that contextualizing recommendations via introductions is an effective strategy for overcoming this challenge and stimulating Taste-Broadening Serendipity.
We identified two distinct mechanisms driving this effect: cognitive elaboration, which offers a reliable and scalable pathway, and transportation, which is more difficult to trigger but has a stronger impact.
These results suggest significant potential for integrating spoken introductions into ``taste-breaker'' playlists on music streaming platforms.
To fully understand the value of such features, future work should examine their long-term impact on sustaining users' listening diversity.
{While manual song curation currently limits scalability, future work could automate introduction generation by using LLMs and knowledge bases.
A TTS setup as detailed in Subsection~\ref{implementation}, can enable the scalable deployment of narrated recommendations across diverse song libraries.}

\section{Ethical and Human Subjects Considerations}
The study protocol was reviewed and approved by the Ethics Committee of our institute. 
Informed consent was obtained from all participants prior to the experiment, as detailed in Section 3.2. 
Participants were explicitly informed of their right to withdraw from the study at any time without penalty.

\begin{acks}
This work was supported by the Research Foundation Flanders (FWO) under grant numbers S006323N, V450225N, and 1158225N.
We are grateful to all our participants for generously contributing their time and insights to this study. We further thank our colleagues, in particular Lien Michiels, Robert Otieno Apiyo, the members of the reading group at SMIT, and the reviewers for their thorough reviews of the manuscript and the many constructive suggestions.
\end{acks}



\end{document}